# Optical Spectra of the High Voltage Erosive Water Discharge

Alexei L. Pirozerski and Sergei E. Emelin

Faculty of Physics, Saint-Petersburg State University, Russia

**Abstract**. In the present paper kinetics of emission spectra of the high voltage erosive water discharge at near ultraviolet and visible spectral ranges has been investigated. Obtained results show a similarity of physical properties of this discharge (and of corresponding plasmoids) to that of some other types of erosional discharges which also result in the formation of dust-gas fireballs.

### 1. Introduction

Recently a special attention was attended to several types of erosional electric discharges giving rise to appearance of long-living plasmoids which have a seeming resemblance with the ball lightning at beginning of their existence, namely, a quasi-spherical form, average size about 10 cm and sufficiently bright luminescence. Among these discharges are HF discharge [1], electric explosion of thin metal diaphragms [2], the high-voltage erosive water discharge (HVEWD) [3,4], non-stationary discharge from 380V power mains (RPMD) [5,6], burning of thin metal wires by electric discharge (MW) [5,7], high-voltage discharge through thin conducting layers (TL) [5,8].

Optical spectra of the discharge's plasma as well as of the plasmoids give an important information about the nature of physical processes occurring therein. The investigations of RPMD, MW and TL discharge's spectra have been performed by us in [6-8], but HVEWD ones remain till recently unexplored. In the present paper for the first time measurements of HVEWD emission spectra kinetics were carried out and a comparative analysis with other erosive discharges is given.

### 2. Experimental setup

The discharge circuit consists of a pulse storage capacitor 1.2 mF x 5kV, an inductance L= 1 mH integrated with pulse ignition transformer (pulse voltage up to 50kV), protecting spark gap (4 mm in length) and HVEWD discharger. Construction of the discharger was similar to described in [4]; a copper cylinder 10 mm in diameter placed inside a ceramic tube was used as the cathode.

For spectral measurements we used a slightly modernized version of the original spectrograph [6]. The spectrograph enables to register the spectral distribution in vertical direction over the cathode at the spectral range 380–650 nm on a camcorder Sony DCR-TRV11E with the frame rate 50 field per second, the maximal spectral resolution being ~1 Å. As a reference source a He-Hg lamp with a reflector was installed behind the discharger. We carried out simultaneous video recording of the objects via another camcorder Sony DCR-HC30E.

### 3. Results and discussion

The HVEWD spectra at different stage of the discharge are shown on the figures below. Indicated times were count off from the discharge ignition with possible error of about ±10 ms. In cases when the discharge spectrum was located on the image sufficiently far from the reference one, a part of the image between them was cut out, on the corresponding figures the spectra being separated by a white line. The reference spectral lines are marked by star (*). The total spectral area was divided into five overlapping ranges. Identification of lines and bands was carried out using tables [9,10].

Fig.1, 2 show the HVEWD spectra at near ultraviolet-violet range (3800–4306 Å). At very beginning of the discharge bright lines of Ca II ion (ionization energy 6.11 eV) at 3968.47 Å (excitation energy $E_{ex}$=3.12 eV) and 3933.67 Å (3.12 eV) are present (Fig.1), then their brightness decreases rapidly (Fig. 2), but they are still visible up to 80 ms after the discharge ignition.  A "line" at ~ 4063 Å is really two lines Cu I 4062.70 Å and 4063.29 Å (6.87 eV); a line near 4023 Å also belongs to copper ( Cu I 4022.65 Å, $E_{ex}$=6.87 eV). These lines are sufficiently bright up to 40 ms, and are still visible at 80 ms. There is a series of lines below 3888 Å marked (1)–(5) at Fig.1,2, which are: (1) – Mg I 3832.3 Å (5.94 eV), (2) – Mg I 3838.29 Å (5.94 eV), (3) – Cu I 3861.74 Å (7.02 eV) and, probably, 3860.47 Å (8.78 eV), (4) – Ca I 3875.81 Å, and 3872.56 Å (5.72 eV), (5) – Ca I 3885.3 Å. These lines are visible up to 60 ms.

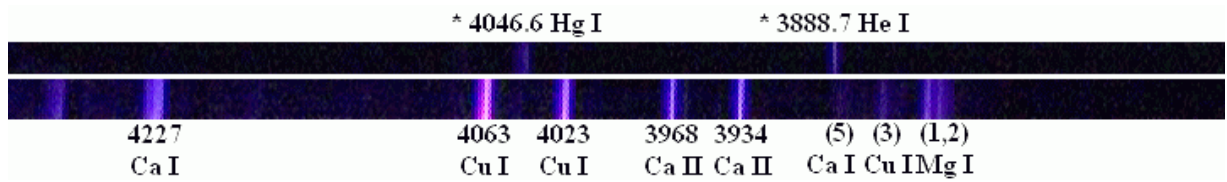

**Fig.1. HVEWD spectrum at UV-V range, 20 ms after the discharge ignition.**

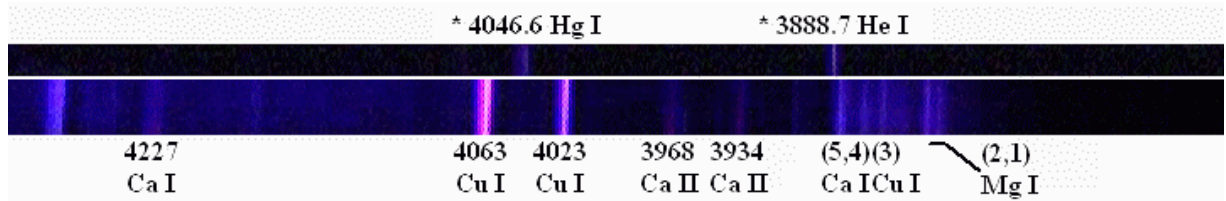

**Fig.2. HVEWD spectrum at UV-V range, 40 ms after the discharge ignition.**

The HVEWD spectra at blue-violet spectral range (4156–4822 Å) at different stages of the discharge are shown on Fig.3 – Fig.5. At the discharge beginning several Cu I lines are visible, which are ( right to left on Fig.3, (1)–(12) ) at : 4275.13 Å ($E_{ex}$=7.74 eV), 4378.20 Å (7.80 eV), 4415.6 Å (7.88 eV), 4480.36 Å (6.55 eV), 4509.39 Å (7.99 eV), 4530.82 Å (6.55 eV), 4539.70 Å (7.88 eV), 4586.95 Å (7.80 eV), 4651.13 Å (7.74 eV), 4674.76 Å (7.80 eV), 4697.49 Å (7.88 eV), 4704.60 Å (7.74 eV).

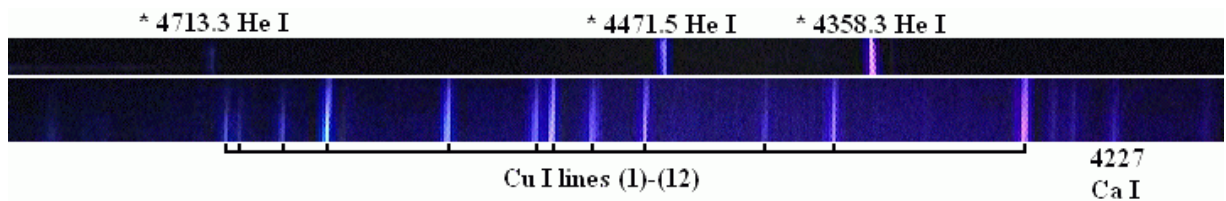

**Fig.3. HVEWD spectrum at V-B range, 20 ms after the discharge ignition.**

To 60 ms after the discharge ignition the spectrum changes drastically (Fig. 4): the Cu I lines have extinguished, the bright line Ca I  4226.73 Å (2.93 eV) is visible (it appears already at 20 ms), as well as several unidentified molecular bands with cants near 4277 Å, 4327 Å, 4334Å, 4338 Å,  4346 Å,  4355 Å,  4363 Å,  4371 Å,  4381 Å,  4389 Å, 4399 Å, 4408 Å, 4419 Å, 4430 Å, 4441 Å, 4451 Å, and also some weak bands at the region 4470–4800 Å.

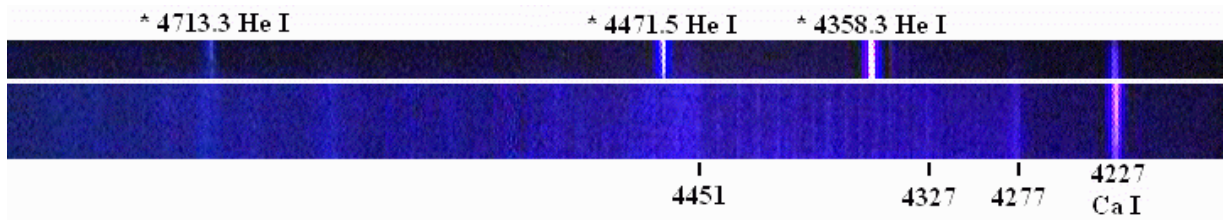
**Fig.4. HVEWD spectrum at V-B range, 60 ms after the discharge ignition.**

At 80 ms line of Ba II ion (ionization energy 5.21 eV) at 4554.04 Å ($E_{ex}$=2.72 eV) and atomic line Ba I at 4604.98 Å (3.81 eV) appear. These two lines are visible up to 140 ms, while the line of Ca I at 4227 Å – up to the current breaking at about 180 ms.

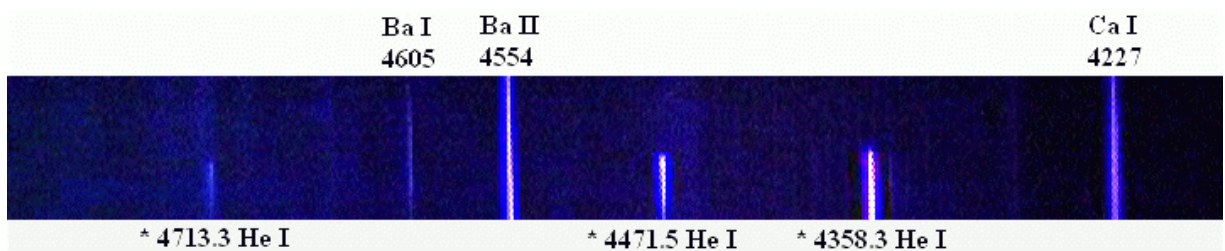
**Fig.5. HVEWD spectrum at V-B range, 100 ms after the discharge ignition.**

The HVEWD spectra at blue-green range (4874–5500 Å) are presented on Fig.6 – Fig.8 below.

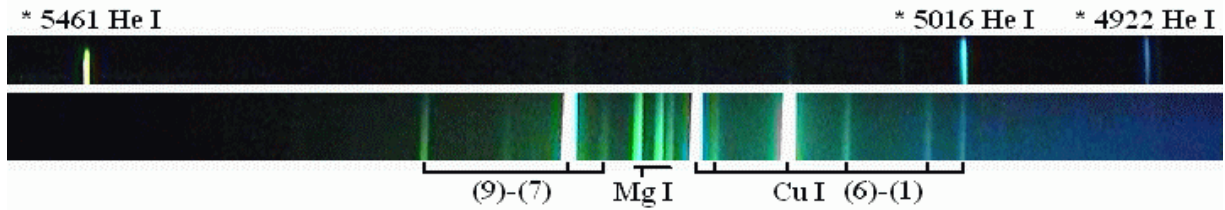
**Fig.6. HVEWD spectrum at the blue-green range, 20 ms after the discharge ignition.**

At the discharge beginning the spectrum is determined by Cu I and Mg I lines. The Cu I lines (right to left on Fig.6, (1)-(9)) are at: 5016.61 Å ($E_{ex}$=7.99 eV), 5034.36 Å (7.99 eV), 5076.17 Å (8.02 eV), 5105.24 Å (3.82 eV) (the rightmost bright line), 5144.12 Å (7.80 eV), 5153.25 Å (6.19 eV) (second bright line), 5200.87 Å (7.80 eV), 5218.20 Å (6.19 eV) (the leftmost bright line). The excitation energy of Mg I lines 5167.33 Å, 5172.68 Å and 5183.61 Å is equal to 5.11 eV.

At 40 ms (Fig.7) the bright line Ba II 4934.09 Å appears, together with several molecular bands with cants near 4949 Å, 4963 Å, 4975 Å, 4986 Å, 4998 Å and 5008 Å attaining maximal brightness at 60 ms. The bands are not identified, but probably are due to BaO.

This spectral range clearly shows the relaxational character of the discharge (excluding its initial stage) and of corresponding plasmoids. Indeed, Cu I lines with excitation energy above 7 eV are visible up to 20 ms after the discharge ignition, the lines 5153.25 Å and 5218.20 Å with $E_{ex}$=6.19 eV – up to ~60 ms, the Mg I lines (5.11 eV) – up to 140 ms, and the Cu I line 5105.24 Å (3.82 eV) is visible up to the current breaking at 180 ms.

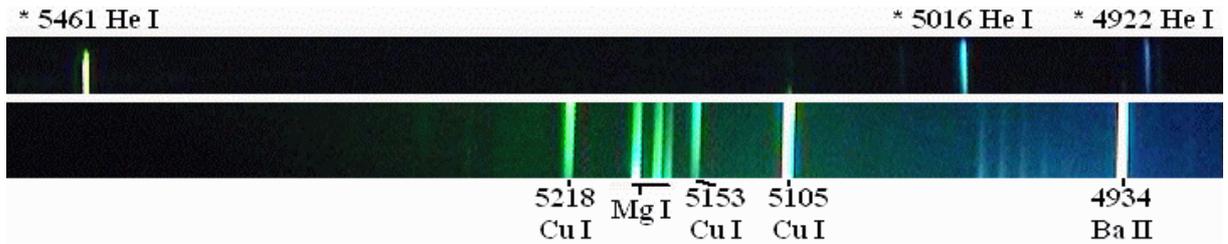
**Fig.7. HVEWD spectrum at the blue-green range, 40 ms after the discharge ignition.**

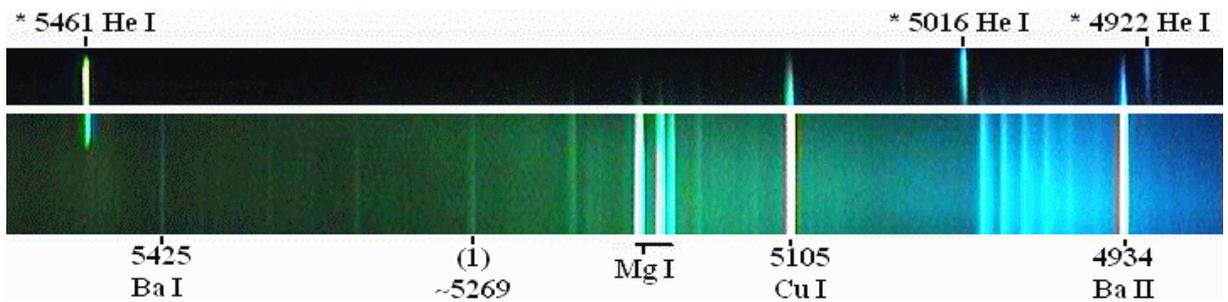
**Fig.8. HVEWD spectrum at the blue-green range, 60 ms after the discharge ignition. (1) is probably the line Ca I 5270.28 Å (4.88 eV) or a band. The excitation energy of the line Ba I 5424.55 Å is 3.81 eV.**

Total energy required for excitation of the Ba II line 4934.09 Å from the atomic ground state is the sum of the ionization energy 5.21 eV and the excitation energy 2.51 eV, so, it is equal to 7.72eV. Nevertheless, the line is visible up to 120 ms. Since after 20 ms in the discharge plasma excitations with energy above 7 eV are practically absent, it is necessary to suppose that injection of barium to the plasma occurs already in the form of ions. Most probably appearance of the barium ions is related with the thermal destruction of barium oxide on the surface of the ceramic tube surrounding the cathode. This assumption explains also the delay in the barium lines appearance – it is necessary for heating the tube to a sufficiently high temperature.

The HVEWD spectra at green-yellow range (5240–5834 Å) are shown on Fig.9, 10. Up to 40 ms after the discharge beginning the spectrum is determined by two Cu I lines – at 5700.24 Å (3.82 eV) and 5782.13 Å (3.79 eV). Their brightness decreases continuously but they are still visible at 140 ms. At ~ 40 ms the line Ba I 5535.48 Å (2.24 eV) appears, its brightness increases up to 100 ms, and the line is sufficiently bright at 160 ms.

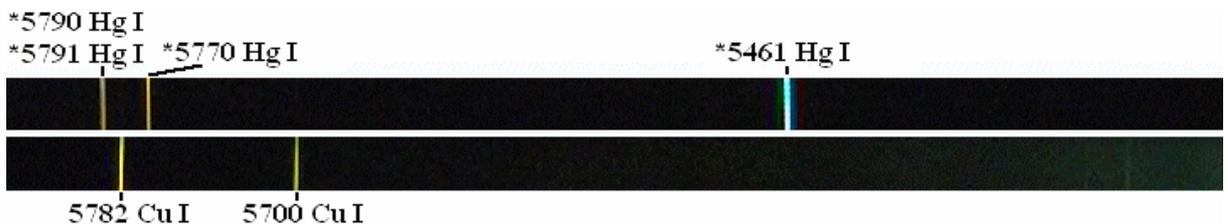
**Fig.9. HVEWD spectrum at the green-yellow range, 40 ms after the discharge ignition.**

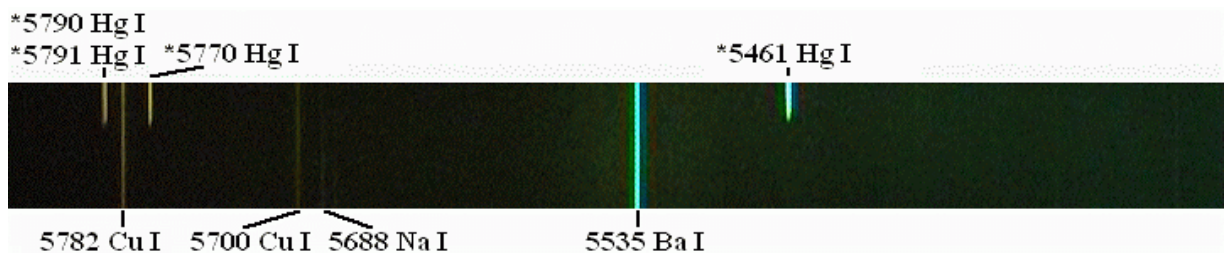
**Fig.10. HVEWD spectrum at the green-yellow range, 100 ms after the discharge ignition.**

In the yellow-red range (5835–6406 Å) from the discharge ignition Na I lines 5889.95 Å (2.11 eV) and 5895.92 Å (2.10 eV) appear (Fig.11) which are visible then during all the discharge and in the afterglow. (It should be noted that in some cases the Na lines were very weak up to ~ 140 ms, then their brightness increased abruptly and remained at high level during 40–60 ms.) At ~20 ms molecular bands of CuO with the cants at 6045.1 Å, 6059.3 Å, 6146.8 Å and 6161.5 Å (Fig.12) appear, which are visible up to the current breaking. Another CuO band with the cant at 6294.0 Å is also present (not shown here).

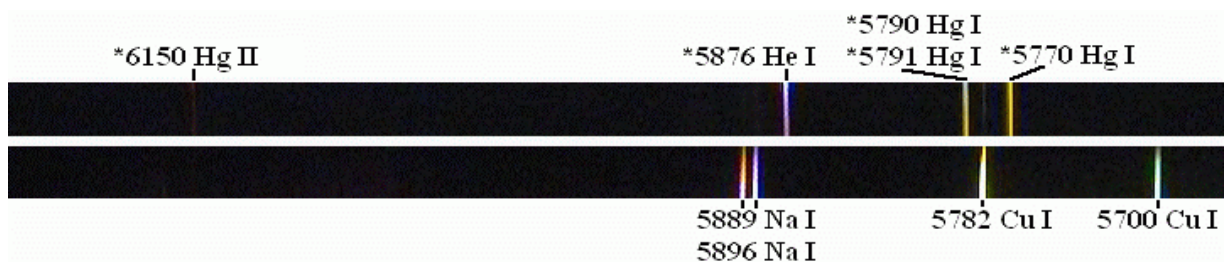
**Fig.11. HVEWD spectrum at the yellow-red range, 20 ms after the discharge ignition.**

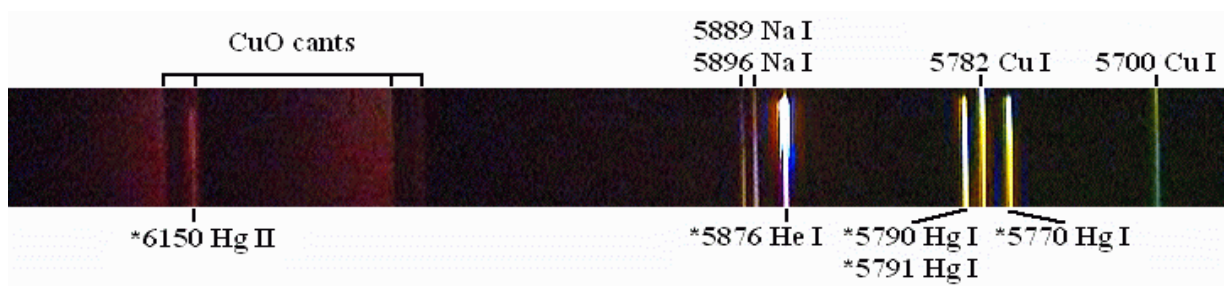
**Fig.12. HVEWD spectrum at the yellow-red range, 80 ms after the discharge ignition.**

Comparing the HVEWD spectra presented above with TL and MW ones [8] we can see that they are very similar while having some minor differences. In particular, in the TL discharge spectrum at the UV-V range for times <30–40 ms bright molecule bands (probably, due to CN) are present below 3888 Å but they are weak or absent in the HWEVD and MW spectra. Next, in the MW and TL spectra there are no Ba I and Ba II lines because of the absence of any ceramic parts in the construction of the corresponding dischargers, but the presence of barium lines in the HVEWD spectra is not internally related with the physical nature of the discharge and only confirms our assertion [6] that the chemical energy of the erosive plasma can be emitted via any available channel with sufficiently low excitation threshold.

So, the spectra of HVEVD, TL, and MW discharges and plasmoids have the following common features:
– the radiation occurs mainly via atomic and ionic lines and via molecular bands, the continuous spectrum is practically absent; this fact, together with the low gas temperature in the discharges, implies non-thermal, nonequilibrium character of the radiation;
– the spectra have relaxational kinetics, i.e. highest energy of available excited states decreases monotonically with time;
– in each case corresponding specific type of the spectra is related with the presence of ones or another chemical elements rather than being directly connected with the kind of active substances accumulating the energy.

### 4. Conclusions

Obtained results support our conclusions [6] that all above mentioned types of the plasmoids have similar physical nature and can be characterized as dust-gas fireballs on the base of chemically-active non-ideal plasma with smallest metal and/or dielectric particles. These objects have only partial similarity with the natural ball lightning; in particular, when disconnected from discharge, they are always relaxing and cannot explode. Their radiation and shape stability related with transfer of chemical energy, stored in the form of radicals, into electron or vibrational excitations. Further advance in experimental modeling of natural ball lightning will require development of new approaches, and one of possible directions is condensed matter based objects.